# Mechanical Acceleration of Capture and Detection Rate of DNA Molecules by Motorizing Bio-Opto-Plasmonic Microsensors

*Jianhe Guo, Kwanoh Kim, Peter Vandeventer, and Donglei (Emma) Fan\**

J. Guo, Prof. D. L. Fan
Materials Science and Engineering Program
Texas Materials Institute
The University of Texas at Austin
Austin, TX 78712, USA
E-mail: dfan@austin.utexas.edu

K. Kim, Prof. D. L. Fan
Department of Mechanical Engineering
The University of Texas at Austin
Austin, TX 78712, USA

P. Vandeventer
Defense Threat Reduction Agency
Fort Belvoir, VA 22060, USA

Abstract: Efficient capture and detection of minute amount of deoxyribonucleic acid (DNA) molecules are pivotal for an array of modern gene technologies which are widely utilized in medical, forensic and defense applications, including DNA extraction, preconcentration, and separation. In this work, we propose a rational mechanism to substantially accelerate the capture and detection efficiency of DNA molecules by actively motorizing designed Raman microsensors. At least 4-fold enhancement has been achieved on the microsensors rotating at 1200 rpm. The process is monitored dynamically and *in-situ* from the biosilica based microsensors, owing to the ultrasensitivity provided by the opto-plasmonic enhancement. The fundamental working mechanism is investigated systematically, and can be attributed to the Nernst diffusion layer thinning effect induced by mechanical motions. This research initiated a new and reliable approach for remarkably enhancing the capture and detection efficiency of biomolecules, which could make far-reaching impact to biosensing, DNA technologies, and microfluidic Total Analysis Systems.

**INTRODUCTION**

Artificial micro/nanomachines made of synthesized micro/nanoscale building blocks have attracted intensive research interest for biomedical applications owing to the versatile motion control and biochemical multifunctionalities.[1-3] An array of applications of artificial micro/-nanomachines have been demonstrated in biomedical research, including drug delivery to single live cells,[4-5] *in vivo* therapeutic delivery in digestive systems,[6-7] tunable drug release,[8-9] assisted fertilization,[10] DNA detection,[11] and RNA sensing inside single cells.[12] These demonstrations showcase micro/nanomachines as unprecedented biomedical nanodevices, exhibiting a changed device scheme from static to active compared to their traditional counterparts. The rapid research progresses of artificial micro/nanomachines are laying an important technological foundation for the emerging micro/nanorobotics on all aspects from materials,[13-14] actuation mechanisms,[15-16] devices,[17-18] to collective operations.[19-20]

In parallel, among an array of modern biotechnologies, deoxyribonucleic acid (DNA) technology is at the core, which has brought revolutionary impact to many aspects of the society, including medical diagnosis, treatment, forensic investigation, and agriculture industry.[21-23] DNA extraction, preconcentration, separation, and purification are one of the most essential processes of the DNA technologies. To accomplish these tasks, the capture of DNA molecules on solid substrates is one of the most promising approaches owing to the facileness in processing and integratiblity with micro/nanofluidic analytic systems.[24-25] A variety of solid particles, including silica,[26-27] polymers,[28] graphene,[29] and gold nanoparticles,[30] have been investigated for this purpose by either electrostatic interaction or covalent bonding of DNA molecules.[31] Among these fascinating nanomaterials, silica nanoparticles have been demonstrated to be one of the most promising candidates due to their advantageous large surface area, unique surface chemistry, biocompatibility, and low cost.[32] Since DNA molecules and silica are both negatively charged and repel each other electrostatically, binding buffers with low pH value and high ionic strength are preferred to reduce the repelling surface charge density and to suppress the electrostatic double layer on the surface of silica, respectively. With these understandings, recent efforts have been focused on developing and optimizing binding buffers to facilitate DNA adsorption on silica, *i.e.* by tuning the pH, ionic strength, and compositions.[33-35] Multiple protocols have been exploited that offer enhanced capture throughput, elution efficiency, and compatibility with downstream applications. However, the time-efficiency of the DNA capture process remains low, particularly for solutions of low molecular concentrations.[36-37] It takes tens of minutes to several hours to fully capture DNA molecules to saturation.[38-40]

How does one substantially improve the DNA capture rate without compromising other important performances? In this research, we exploit and test a new device of rotary micromotor-sensor unibody, which can offer prominently enhanced capture and detection efficiency of biomolecules, *e.g.* DNA, without compromised performances in sensitivity, a Holy Grail in biodetection. The rotary micromotor-sensors are made of periodically nanoporous diatom frustules, which provide hybrid opto-plasmonic modes when coupled with silver (Ag) nanoparticles, as determined by both experiments and simulations. They can substantially promote detection sensitivity of the label-free Raman Spectroscopy for monitoring the dynamic capture process of DNA molecules *in situ* and in quasi-real-time. Next, assisted with external magnetic fields, we facilely motorize and assemble the microsensors into microfluidic wells and channels, and rotate them with tunable speeds up to 1200 rpm. The Raman spectra obtained from the microsensors unveil the dynamic capture process, which is clearly accelerated by the mechanical rotation. The higher the rotation speed, the higher the capture rate. We investigate fundamental mechanisms systematically, which can be attributed to the Nernst diffusion layer thinning effect resulting from the mechanical rotation. This effect is confirmed in enhancing the capture and detection speed of biomolecules on micro/nanoscale sensors for the first time. The reported working mechanism and rational approach could open an array of opportunities for DNA technologies, biosensing, and single-cell analysis.

**RESULTS**

**Design and Fabrication of Magnetically and Plasmonically Active Biosilica Micromotor-Sensors**

A bio-opto-plasmonic microsensor is designed consisting of a diatom frustule with uniformly-distributed plasmonic Ag nanoparticles on the surface and a magnetic thin film deposited on one side [Figure 1(a)]. Diatoms are one of the most common and widespread types of photosynthetic microalgae, living abundantly in marine ecosystems. Frustules are the cell walls of diatoms made of silica with ordered arrays of nanopatterns [Figure 1(b-c)].[41-42] These naturally existing and abundantly available biomaterials could be beneficial for low-cost micro/nano-manufacturing. In particular, the mechanical strength and chemical stability of these diatom frustules suggest potential applications for micro/nanomechanical devices, such as rotary micromotors.[43] The silica chemistry, large surface area, and low cost make them excellent candidates for DNA capture. The periodic nanoporous structures further support an effective management of light, *i.e.* functioning as photonic crystals that resonate with light for substantial enhancement of optical signals of molecules for detection.[44-45] Here, we refined and obtained diatom frustules from low-cost commercial diatomaceous earth powder after processing through the dispersion, sonication, filtration and calcination treatments (experimental details are provided in Supporting Information S1).[43] The dispersion, sonication, and filtration processes take out impurities and broken pieces of frustules; the calcination process burns off organic residues, delivering purified cylindrical diatom frustules with diameters of 8 to 16 μm and lengths of 20 to 35 μm [Figure 1(b)]. Next, a bilayer nickel (Ni)/gold (Au) thin film was deposited on one side of the diatom frustules for magnetic manipulation in suspension. The ferromagnetic response of Ni to an external magnetic field aligns diatom frustules in the direction of magnetic fields and transports them along the gradient of magnetic fields. Finally, plasmonic Ag nanoparticles were grown uniformly on the surface of diatom frustules for surface-enhanced Raman spectroscopy (SERS) detection [Figure 1(d)].[9, 46] The plasmonic Ag nanoparticles are densely distributed on the surface of diatom frustules with an average diameter of 23.38±3.30 nm and junctions ranging from 0.5 nm to 5 nm (Supporting Information Figure S1).

**Assembling and Rotating Micromotor-Sensors in Microfluidic Wells**

We applied rotating magnetic fields with a commercial magnetic stirrer arranged on the top of the micromotor-sensors in microfluidics as shown in Figure 2(a). This setup is based on the widely used magnetic stirrer available in most chemistry and biology labs. The top-arrangement of the magnetic stirrer allows the generation of external magnetic forces on microsensors in the upward direction, proportional to the field gradient as given by $\boldsymbol{F_m} = (\boldsymbol{m} \cdot \nabla)\boldsymbol{B}$, where $\boldsymbol{m}$ is the magnetic moment of microsensors.[47] The magnetic force balances with the gravitational force and effectively reduce the hindering force on a microsensor from the substrate. As a result, it ensures a stable and durable rotational operation of the microsensors.[48] This is particularly desirable for capture of DNA molecules, the binding buffers of which with low-pH and high ionic strength result in lower electrostatic repulsion and stronger wall effect that increase hydrodynamic drags, compared to those in low ionic suspensions.[49]

When the magnetic stirrer is set to rotate, the magnetic torque, determined by $\boldsymbol{\tau_m} = \boldsymbol{m} \times \boldsymbol{B}$,[50] compels the rotation of multiple microsensors at a synchronous speed. The speed can be tuned up to 1200 rpm [Supporting Information Figure S2]. During the continuous rotation, microsensors self-assembled into microfluidic wells and channels [Figure 2(b-c) and Video S1]. Every micromotor-sensor takes an individual microwell. When a microsensor encounters a microwell occupied by a rotating microsensor on its moving trajectory, it detours around and moves forward as shown in Video S1. The observed one-on-one assembling of microsensors in individual microwells is highly interesting for assembly of devices in microfluidics. To understand the effect, we carry out COMSOL simulations to determine fluidic fields generated by two neighboring rotating microsensors as shown in Figure 2(e-h). It is found that each microsensor produces a rotational liquid flow and experiences a flow with a gradient of shear produced by the other microsensor. The background shear rate has a dependence of $\gamma(d,\theta) \propto v(d,\theta)/D \propto P(\theta)\omega l^3/d^3$,[51] where $d$ and $\theta$ are the distance between the center of microsensors and their rotating angle, respectively, $v(d,\theta)$ is the fluidic velocity at the location of microsensor, $l$ is the length of the microsensor, $\omega$ is the angular rotation speed, and $P(\theta)$ is a parameter depending on $\theta$. Due to the inertial effect, each microsensor experiences a hydrodynamic repulsive force from the other microsensor, given by $F_r \propto \eta \omega l^2 Re$,[51] where $\eta$ is the viscosity and $Re$ is the Reynolds number. The Reynolds number $Re$ is calculated based on the shear rate ($\gamma$), $Re = \rho \gamma l^2/\eta$, where $\rho$ is the density of medium. Therefore, it can be found $F_r \propto P(\theta)\rho \omega^2 l^7/d^3$. The generated flows always produce a hydrodynamic repulsive force between the two microsensors at all rotary configurations, *i.e.* at 0˚, 45˚, 90˚, and 135˚. The higher the rotating speed, the higher the repulsion force. Note that the hydrodynamic repulsive force ($F_r \propto 1/d^3$) increases much faster than the magnetic attraction force ($F_m \propto 1/d^2$) when two rotating microsensors approach each other. This explains the experimentally observed repulsion of synchronously rotating microsensors, no matter if they approach each other end to end (magnetically attracting) or side by side (magnetically repelling). This facile assembling approach could be potentially adopted by a variety of microfluidic devices and systems.[52-53]

**SERS Characterization of the Opto-plasmonic Micromotor-Sensors and Numerical Modeling**

Next, we characterized the Raman sensitivity of the microsensors before studying their interaction with DNA molecules during mechanical rotation. Conventionally, the capture dynamics of DNA molecules by solid substrates is determined by measuring the concentration of DNA molecules remaining in solutions at fixed time intervals with Ultraviolet (UV) absorbance spectroscopy.[39] The most advanced UV absorbance technique requires at least microliter sized samples for accurate measurement. It is also labor-intensive and time-consuming in revealing capture dynamics of molecules. Here, we employ SERS spectroscopy to directly detect DNA molecules captured on the silica surface. Since the plasmonic hotspots for SERS detection present in the vicinity of Ag nanoparticles and their narrow junctions, only the DNA molecules attached on the surface of the microsensors can be detected. It has been demonstrated that Raman signals of molecules can be enhanced up to ~ $10^9$ to $10^{10}$ on the surface of plasmonic substrates.[46, 54] With such a high surface enhancement, the background signals of molecules in suspension are greatly minimized. As a result, the dynamic attachment dynamics of molecules on the plasmonic surfaces can be unveiled *in-situ* at a high speed.

Our study begins with quantitative determination of SERS enhancement of the biosilica-based opto-plasmonic microsensors by carrying out both experiments and numerical simulation. The Raman signals of DNA molecules were collected as shown in Figure 3(a) after incubation in DNA solutions ranging from 80 nM to 4 μM for 2 hours. At the lowest detectable concentration of 80 nM, the signal-to-noise ratios are 4.21 and 5.79 at 724 cm$^{-1}$ and 1319 cm$^{-1}$ (both attributed to the vibration of nucleobase adenine), respectively.[55-56] The Raman enhancement is uniform along the entire length of the microsensors [Figure 3(b)]. Compared to plasmonic Ag nanoparticles synthesized at the same condition on a flat glass substrate, the Raman intensity obtained from plasmonic biosilica microsensors is further improved by 3.66 ± 0.31 times as shown by all major Raman peaks of nucleobases and deoxyribose [Figure 3(c)], *i.e.* 724 cm$^{-1}$ (adenine), 957 cm$^{-1}$ (deoxyribose moiety), 1235 cm$^{-1}$ (thymine and cytosine), 1319 cm$^{-1}$ (adenine) and 1391 cm$^{-1}$ (guanine).[55-56] The further improvement of Raman signals can be attributed to the resonance of localized surface plasmons with the photonic nanostructures of the diatom frustule as shown by the numerical simulation in Figure 3(d). The maximum electric-field intensity of a pair of Ag nanoparticles placed on a diatom frustule with ordered photonic nanopores is as high as 1.98 times of that obtained from the same dimer on a flat glass substrate.

The simulation details are provided in the Supporting Information. This substantial enhancement of Raman sensitivity from the unique diatom frustule based Raman microsensors allows sensitive detection of DNA at low concentration in a label-free fashion. This advantageous sensitivity readily facilitates the monitoring of DNA capture dynamics during rotation of the microsensors in quasi-real time.

**Mechanically Accelerating the Capture and Detection of DNA Molecules**

The microsensors assembled in microwells rotate at speeds tunable up to 1200 rpm. The dynamics of DNA capture process is monitored by Raman spectroscopy every 10 seconds continuously for 30 minutes. The Raman intensity of DNA molecules increases monotonically until essentially reaching a saturation equilibrium on the microsensors as shown in Figure 4. Reproducibly, the mechanical rotation accelerates DNA capture. For instance, the required time to reach 95% of the capture equilibrium of DNA is reduced substantially with mechanical rotation, from ~ 28 minutes on a static microsensor, to ~ 7 minutes on a rotating microsensor at 1200 rpm [Figure 5(a)]. Even at a speed of 300 rpm, the capturing process only requires one-third of the time compared to a static microsensor. Although the acceleration effect with mechanical rotation has a sub-linear dependence on rotation speed [Figure 5(c)], the 4-fold improvement obtained at 1200 rpm can already significantly enhance the efficiency of DNA capture, which has great potential benefits to a variety of DNA technologies.

With in-depth analysis, we modeled the capturing kinetics of DNA on microsensors with the Nernst diffusion theory and Langmuir adsorption theory.[57-58] As shown in Figure 5(b), three steps are involved in the DNA capture process from the bulk solution to the silica surface: the convection-dominated transport of DNA to the outer boundary of the stationary Nernst diffusion layer, the diffusion-dominated transport of DNA inside the Nernst diffusion layer, and the final adsorption of DNA molecules on the silica surface.[59] As suggested by the Nernst diffusion theory, the diffusion layer is the stationary region between the solid surface and mobile liquid, where the transport of molecules only occurs in the form of passive diffusion driven by the concentration gradient.[58] Due to mechanical rotation of the microsensors and the resulted flow convection of liquid, the DNA concentration is considered constant at the outer boundary of the diffusion layer. As a result, the overall capturing kinetics is dominated by the sequential diffusion process in the Nernst diffusion layer and adsorption process on the silica surface.

In the process of molecule diffusion through the Nernst diffusion layer to the solid surface, the concentration gradient between its outer-layer boundary ($C_{outer}$) and inner boundary ($C_{inner}$) drives the diffusion process of DNA, the dynamics of which follows Fick's laws below:[58]

$$C_{inner} = C_{outer} - C_{outer} \cdot e^{-k_d t}, \quad (1)$$

where $C_{outer}$ is same as the bulk concentration ($C_0$), and $k_d$ is the diffusion rate. As shown in the calculation in the Supporting Information, $k_d$ is inversely proportional to the thickness of Nernst diffusion layer ($\delta$), $k_d \sim \delta^{-1}$. According to the Levich equation,[60] due to the shear flow on the solid surface of rotating objects, the thickness of the Nernst diffusion layer is inversely proportional to the square root of the rotation speed ($\omega$): $\delta \propto \omega^{-0.5}$, which is also known as the fluidic boundary layer effect. Therefore, $k_d \propto \omega^{0.5}$, qualitatively proving that the diffusion kinetics of DNA molecules to the surface of the microsensor can be actively controlled by the mechanical motions of the microsensors. The higher the rotation speed ($\omega$), the higher the diffusion rate ($k_d$), with a square root dependence.

After diffusing through the Nernst diffusion layer, molecules are absorbed on the solid surface, which is another process that can control the overall kinetics. This absorption process can be modeled by the Langmuir adsorption theory, where the concentration of molecules adsorbed on a solid surface at equilibrium ($C_{s,eq}$) is a function of its concentration in the bulk solution ($C_0$), adsorption equilibrium constant ($K_{eq}$), and the maximum adsorption capacity of the material ($C_{s,max}$), given by:[57]

$$C_{s,eq} = C_{s,max} \frac{K_{eq}C_0}{1+K_{eq}C_0}. \quad (2)$$

Therefore, the experimentally determined linear dependence of intensity of Raman signals and concentration of DNA in Figure S3 in the Supporting Information, indicates the concentrations of DNA are much below the saturation limit of the microsensors in the tested concentration range. The time dependent concentration of DNA molecules adsorbed on the surface ($C_s$) is given by the Lagergren equation as below:[57]

$$C_s = C_{s,eq} - C_{s,eq} \cdot e^{-k_a t}. \quad (3)$$

Note that, to the first order of approximation, the adsorption rate ($k_a$) is considered independent on the flows generated by the rotating microsensors. It can be found that both the diffusion process (Equation 1) and adsorption process (Equation 3) are governed by a equation with a same form as below:

$$C_s = C_0 - C_0 \cdot e^{-kt}, \quad (4)$$

where $k$ is the process controlling rate. If diffusion is the limiting process in the overall capture kinetic ($k_d < k_a$), $k$ is taken as $k_d$; by the same token, if adsorption is the limiting process ($k_a < k_d$), $k$ is taken as $k_a$. When both processes play close roles, $k$ is governed by both $k_a$ and $k_d$.

In our experiments, we obtain DNA capture rates ($k$) at different rotation speeds by fitting results of Raman Intensity of DNA versus time in Figure 4 with Equation 4. The obtained capture rates ($k$) versus rotation speed ($\omega$) are shown in a log-log plot in Figure 5(c). It is found that the DNA capture rate ($k$) monotonically increases with rotation speed ($\omega$) from 60 rpm to 300 rpm, with a power-law dependence of 0.49 ± 0.02, well agreeing with that predicted by the Nernst diffusion layer thinning effect, where $k_d \propto \omega^{0.5}$. It also indicates that the diffusion of DNA through the Nernst diffusion layer is the limiting step in the overall capture process at this rotation speed range. When the rotation speed is greater than 600 rpm, the value of $k$ starts to deviate from the linear log fitting of $k_d$ versus $\omega$ with a slope of 0.5, suggesting the competition of the adsorption in governing the overall DNA capture kinetics. Overall all the experimental observation can be well understood by our modeling: at a low rotation speed of a microsensor, diffusion of DNA through the Nernst diffusion layer to the surface of microsensor is the kinetic limiting step ($k_d < k_a$); $k$ corresponds to $k_d$ with an inverse-square-root dependence on $\omega$. At a high rotation speed of a microsensor, *i.e.* greater than 630 rpm as observed, the enhanced diffusion rate ($k_d$) due to the rotation of the microsensor starts to become comparable with the adsorption rate $k_a$. Both $k_d$ and $k_a$ contribute to the observed capture kinetics. As a result, the experimentally determined dependence of $k$ on $\omega$ deviates from the inverse-square-root relationship.

To further understand the kinetic limiting steps presented in the capture process of DNA molecules, we studied the capture dynamics of DNAs at different temperatures. Since the investigation of DNA capture dynamics were carried out at the temperature range of 5-45 °C assisted with constant temperature incubators, *i.e.* a refrigerator and a thermostatic water bath, we utilized UV spectroscopy to determine the concentrations of DNA molecules during the adsorption process. It is known that the adsorption of DNA on the silica surface in a binding buffer with pH lower than 7 is an exothermic process.[38] Therefore, the lower the temperature, the higher the adsorption rate. In comparison, the diffusion process is governed by the Arrhenius equation, where a higher temperature facilitates the diffusion rate. Clearly, these two effects have countering dependence on temperature. Therefore, by examining dependence of DNA capture rate on temperature, we can further confirm the dominating rate limiting process. As shown in Figure S4, experimentally, we found the concentration of DNA molecules in bulk solution decreases with time at all tested temperatures, indicating the successful detection of the DNA adsorption process. Furthermore, we found that the adsorption rate of DNA monotonically increases with temperature. This dependence supports our earlier finding that it is the diffusion of DNA molecules to the surface of microsensor that dominates the overall capture kinetics. Therefore, to enhance the capture efficiency of DNA molecules by the silica microsensors in the interesting temperature and concentration ranges in our tests, the key is to improve the diffusion rate of DNA molecules, which can be effectively achieved by flows generated with mechanical rotation of the microsensors.

Next, with the above understanding, we apply the motorization strategy to realize high-speed capture of low-concentration DNA and its detection with Raman spectroscopy. Regardless of the detection mechanisms as optical, mechanical, or electrical, the high sensitivity comes at the expense of an adversely long detection time for low-concentration analytes.[61-62] This challenge originates from the slow diffusion of analytes to the surfaces of micro/nanosensors, which can be understood by the low probability for low-concentration molecules to attach to micro/nanosensors. Here, we demonstrate the first successful enhancement of detection efficiency of low-concentration molecules by mechanically rotating a micro/nanosensor. Samples with a DNA concentration of 80 nM are tested on both static and rotating microsensors. The speed of rotation is 630 rpm. Figure 6 presents the obtained time-dependent Raman spectra of DNA molecules. The Raman peak at ~780 cm$^{-1}$ from silica is constant during the detection and is used as a reference for comparison with DNA signals. On the static microsensor, the Raman signal of DNA at 724 cm$^{-1}$ was hardly observable in the first 3 minutes, and became gradually clear at ~ 10 minutes during the DNA detection process. From the rotating microsensor, however, it took only 3 minutes to exhibit the Raman signal of DNA molecules with an intensity similar to that obtained at around 10 minutes from the static microsensor. The improvement is ~ 3 times in detection speed, demonstrating the feasibility in obtaining high-speed detection of low-concentration molecules with motorized micro/nanosensors, via thinning the Nernst diffusion layer to accelerate the diffusion of analytes for detection.

**CONCLUTION**

In summary, an innovative, effective, and robust mechanism for enhancing the capture rate of DNA molecules is reported. The opto-plasmonic microsensors made of diatom frustules provide biocompatible silica for DNA capture, and periodic nanopore arrays coupled with plasmonic Ag hotspots for sensitive and quasi-real-time SERS detection. The microsensors are transported and self-assembled in individual microfluidic wells or channels for capturing and detecting DNA molecules during high-speed rotation. The mechanical rotation accelerates the capture process for both high and low-concentration DNA molecules, *i.e.* 4 times at a speed of 1200 rpm. The fundamental mechanism is modeled and understood by the Nernst diffusion and Langmuir adsorption theories, and the Nernst diffusion layer thinning effect, with quantitative agreement. The new concept presented in this work could inspire many new opportunities, paradigms, and devices for high-speed biochemical enrichment, reaction, and sensing, relevant to biodetection, DNA technologies, and microfluidic Total Analysis Systems.

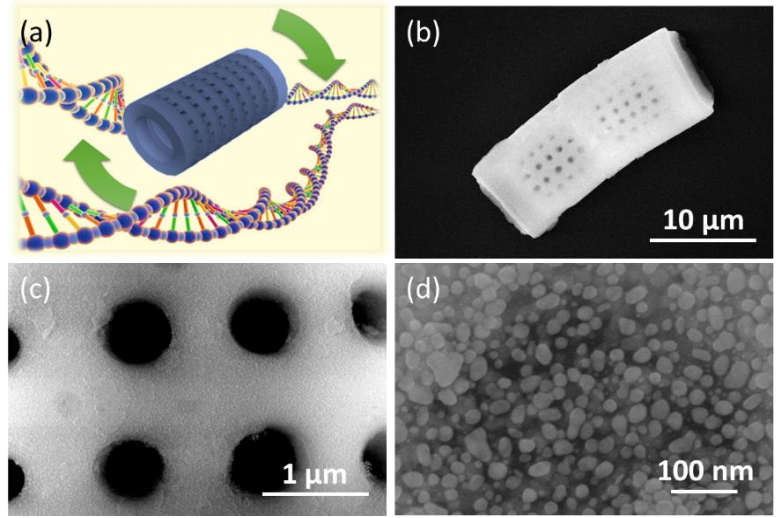

**Figure 1**. (a) Schematic of a rotating biosilica opto-plasmonic microsensor in DNA solutions. (b-d) Scanning electron microscopy (SEM) images of diatom frustules with periodic nanopores, and (d) Ag nanoparticles synthesized on the surface.

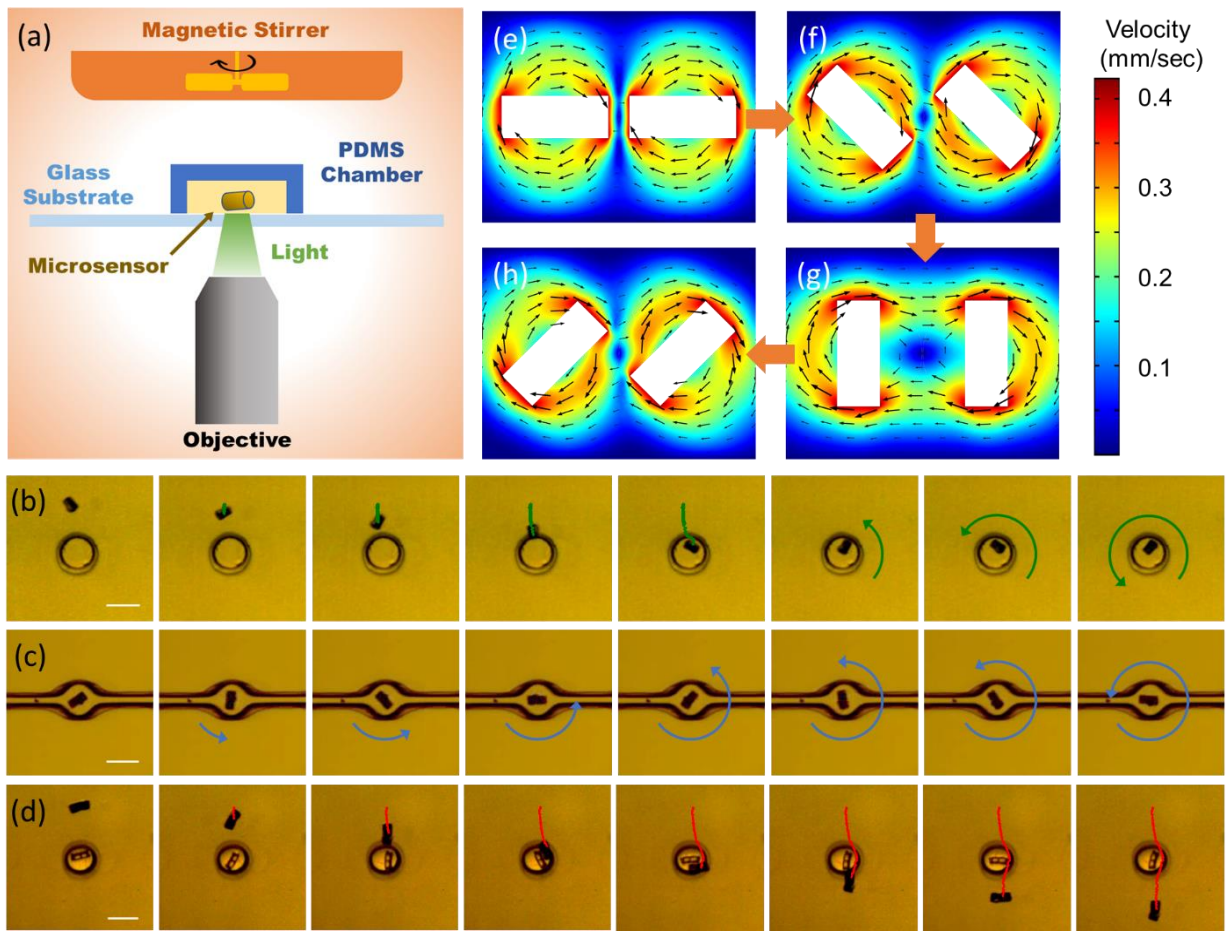

**Figure 2**. (a) Schematic of experimental setup for detection of DNA molecules from the motorized microsensors. (b) A microsensor rotating and assembling into a microwell. First 5 snapshots are taken every 1 second. The following snapshots are taken every 0.05 second. (c) A microsensor rotating in a microfluidic channel, snapshots taken at every 0.025 second. (d) A microsensor rotating at 300 rpm detoured around an occupied microwell. Snapshots taken every second. Scalebar in (b-d) is 30 μm. (e-f) Simulation of the fluidic fields around two neighboring microsensors rotating synchronously at (e) 0° (end to end), (f) 45°, (g) 90° (side by side), and (h) 135°.

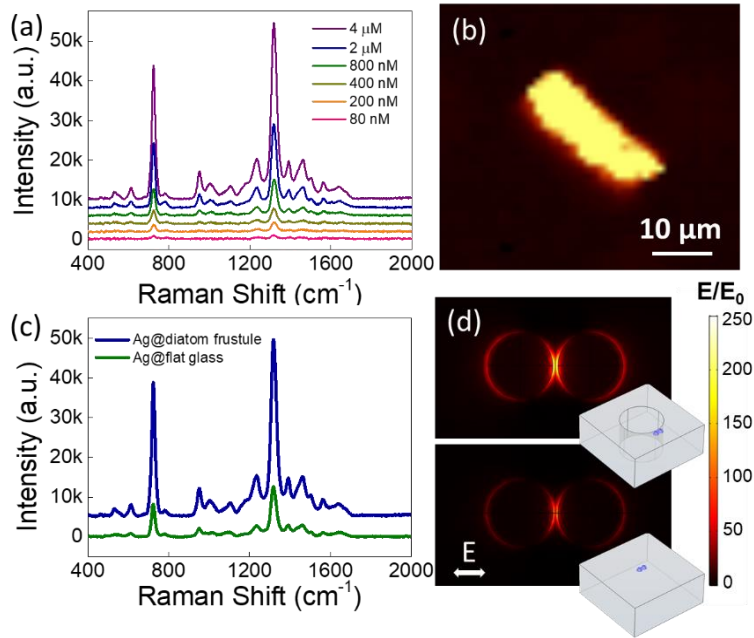

**Figure 3**. (a) SERS spectra of DNA molecules detected from 80 nM to 4 µM. (b) Raman mapping of DNA molecules (4 µM, 724 cm$^{-1}$) on a microsensor after 2-hour incubation. (c) SERS spectra of 4 µM DNA on a microsensor (blue) and on a Ag-nanoparticle coated flat glass substrate (green), respectively. (d) Simulations of electric field distributions around a pair of Ag nanoparticles placed on a diatom frustule substrate (top) and a flat glass substrate (bottom), respectively. Insets are the schematics of the simulated structures with Ag nanoparticles. The incident laser is polarized horizontally.

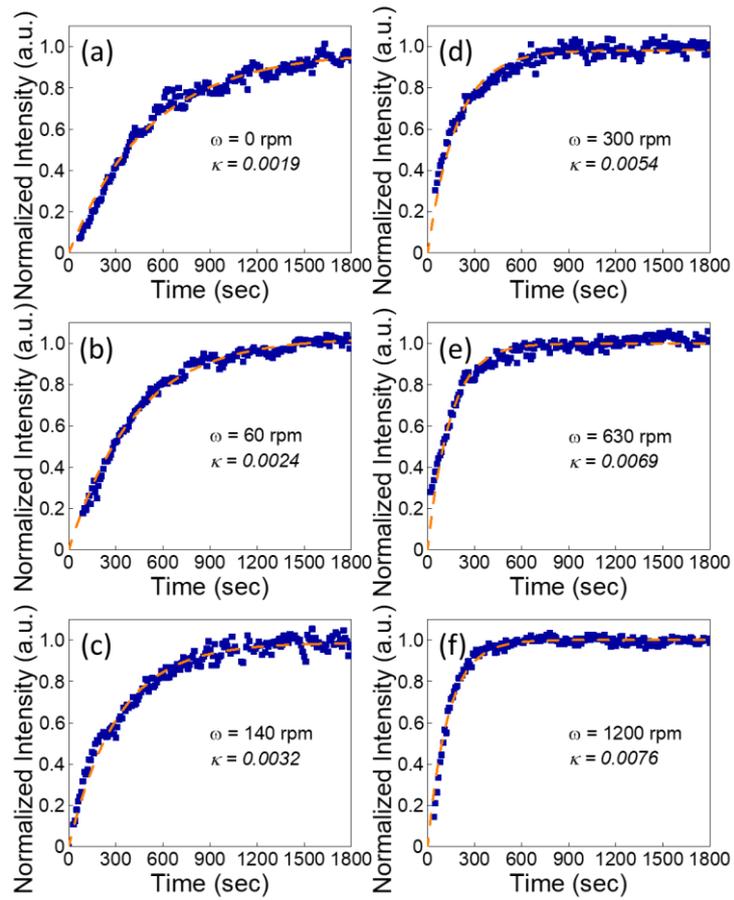

**Figure 4**. Dynamics of the Capture process of DNA molecules revealed by SERS on (a) a static microsensor and (b-f) a rotating microsensor at (b) 60 rpm, (c) 140 rpm, (d) 300 rpm (e) 630 rpm, and (f) 1200 rpm.

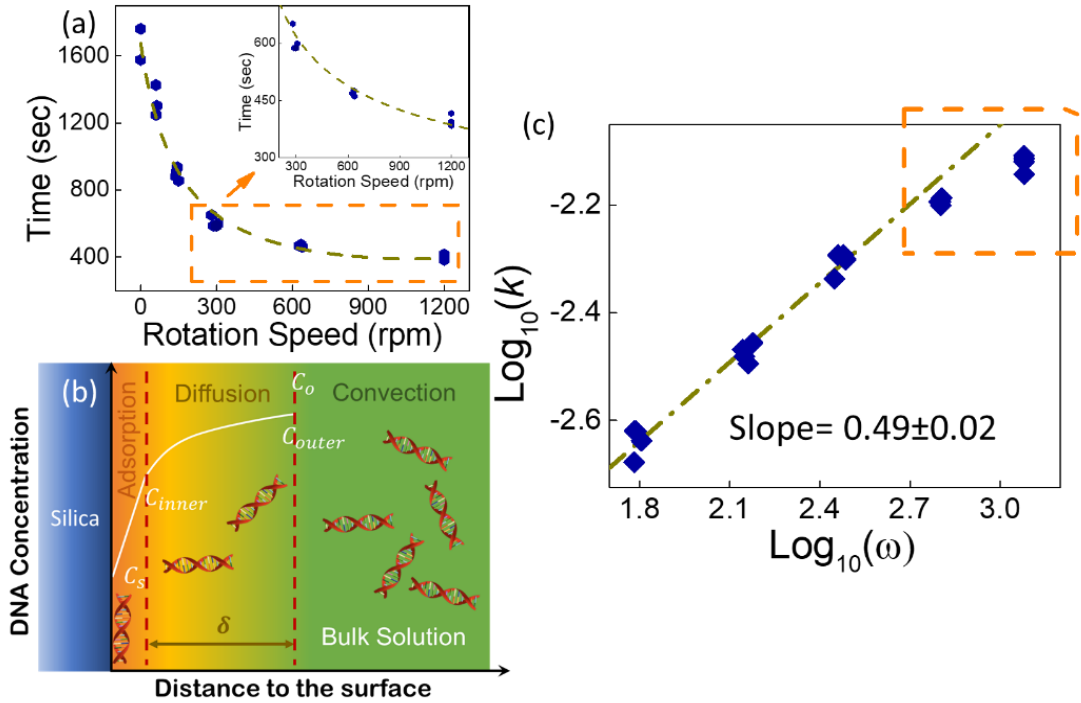

**Figure 5**. (a) Time needed to attain 95% of the saturation of DNA capture versus rotation speed of microsensors. (b) Schematic model: the DNA capture dynamics enhanced by mechanical rotation, including convection, diffusion and adsorption. (c) Log-log plot of capture rate ($k$) versus rotation speed ($\omega$ in rpm) of microsensors. The linear fit of results from 60 rpm to 300 rpm shows a slope of 0.49.

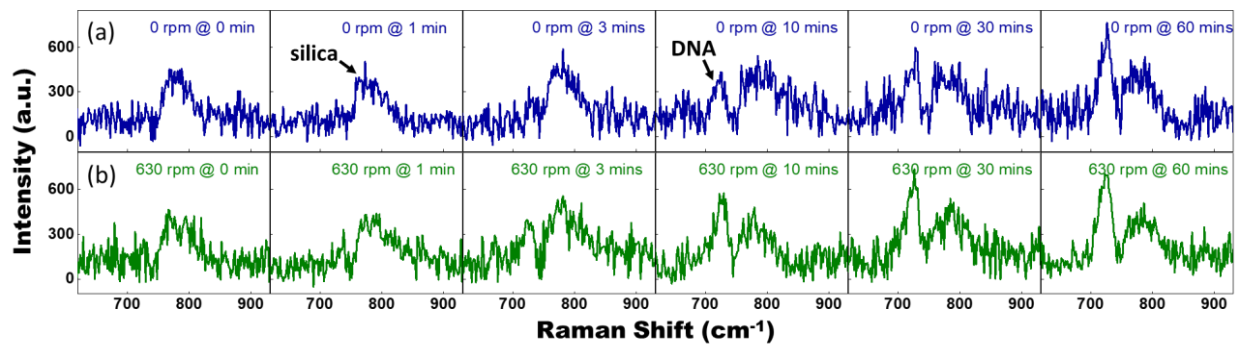

**Figure 6**. Raman spectra of DNA molecules (80 nM) measured at fixed time intervals for an hour on (a) a static (in blue) and (b) a rotating microsensor with speed of 630 rpm (in green).

23